\documentclass[aps,prl,twocolumn,superscriptaddress,showpacs]{revtex4-2}
\usepackage[T1]{fontenc}
\usepackage[utf8]{inputenc}
\usepackage[a4paper]{geometry}
\usepackage{amsmath}
\usepackage{relsize}
\usepackage{amssymb}
\usepackage{amsfonts}
\usepackage{graphicx}
\usepackage[export]{adjustbox}
\usepackage{subfigure}
\usepackage{mathtools}
\usepackage{csquotes}
\usepackage{comment}
\usepackage{xcolor}
\usepackage{longtable, multirow}
\usepackage{pifont,array}
\usepackage{hyperref}
\usepackage{tikz}
\usetikzlibrary{decorations.markings}
\newcommand\underrel[3][]{\mathrel{\mathop{#3}\limits_{%
      \ifx c#1\relax\mathclap{#2}\else#2\fi}}}
\newcommand{\im}{\mathrm{i}}

\renewcommand{\rm}[1]{\mathrm{#1}}
\newcolumntype{"}{!{\vrule width 1pt}}
\newcommand{\phs}{\mathcal{P}}
\newcommand{\eul}{\mathrm{e}}

\definecolor{ffzztt}{rgb}{1,0.6,0.2}
\definecolor{yqqqqq}{rgb}{0.5019607843137255,0,0}
\definecolor{ffzzqq}{rgb}{1,0.6,0}
\definecolor{qqwwzz}{rgb}{0,0.4,0.6}
\begin{document}

\title{Softening of Majorana edge states by long-range couplings}

\author{Alessandro Tarantola}
\affiliation{Institut f\"ur Theoretische Physik, ETH Z\"urich,8093 Z\"urich, Switzerland}

\author{Nicol\`o Defenu}
\affiliation{Institut f\"ur Theoretische Physik, ETH Z\"urich,8093 Z\"urich, Switzerland}

%\preprint{}

\begin{abstract}
The inclusion of long-range couplings in the Kitaev chain is shown to modify the universal scaling of topological states close to the critical point. By means of the scattering approach, we prove that the Majorana states \textit{soften}, becoming increasingly delocalised at a universal rate which is only determined by the interaction range. This edge mechanism can be related to a change in the value of the bulk topological index at criticality, upon careful redefinition of the latter. The critical point turns out to be topologically akin to the trivial phase rather than interpolating between the two phases. Our treatment moreover showcases how various topological aspects of quantum models can be investigated analytically. \end{abstract}

\maketitle

%\tableofcontents

\section{Introduction} \label{sec:Intro}

Efficient quantum computing is probably the primary goal of modern physics research and the race for quantum advantage involves research groups all around the globe\,\cite{higgins2007fundamental,carlo2009demonstration,monroe2013trapped,debnath2016small,barends2014universal,ofek2016extending,arute2019quantum}. The first spark to this intense research activity came from the formulation of Shor's algorithm for prime factorization\,\cite{shor1994algorithms,shor1997polynomial}, which was followed by a large number of influential theoretical proposals\,\cite{cirac1995quantum,beckman1996efficient,lloyd1996universal,wiebe2012quantum,biamonte2017quantum}. Nowadays, state of the art quantum simulators include ensembles of superconducting qubits\,\cite{blais2021circuit}, trapped ions\,\cite{monroe2021programmable}, cold atoms confined in optical cavities\,\cite{ritsch2013cold,mivehvar2021cavity} and Rydberg atom experiments\,\cite{adams2019rydberg,chomaz2022dipolar}. Interestingly, most of these quantum platforms feature long-range power-law decaying interactions, which are hence becoming an essential ingredient of modern quantum simulation\,\cite{defenu2021quantum}.

Although not yet realized in experiments, topological quantum computation represents a promising route to fault tolerance\,\cite{kitaev2003fault}. As a consequence, recent years have also witnessed outstanding efforts in the experimental and theoretical study of topological matter \cite{TKNN,Hatsugai93,KitaevTable,AKLT,sau2010generic,mong2014universal,biao2018topological,NadjPerge2014}, see Refs.\,\cite{nayak2008non,stern2013topological} for a review. A promiment model exhibiting topological nature is surely the Kitaev chain (KC) \cite{KitaevChain}, a 1D superconductor with nearest neighbor (NN) hopping and pairing. Its most striking feature is the presence of unpaired Majorana zero-modes at the edges of the sample. Such modes have been proposed as ideal \enquote{topological qubit} candidates, thereby igniting extreme interest in the quantum information and computation communities\,\cite{MajoQC1,MajoQC2,MajoQC3,Alicea2011,Alicea2012,Aasen2016}. Not only is this model an ideal theoretical playground, but increasingly sophisticated experimental realizations are seeing the light. For example, \cite{Dvir23} uses quantum dots coupled via electron tunneling and crossed Andreev reflection, whereas \cite{Rancic22} implements the model on the three noisy qubits of a publicly available quantum computer. Superconductor-semiconductor (SC-SM) nanowires, specifically devised for experimental detection of zero-modes, have also been shown to map to such model \cite{Pan2021}.

Long-range effects and topology happily marry in the Kitaev chain, upon long-range (LR) extension of the latter by endowing the hopping ($j$) and pairing ($\Delta$) terms with a dependence on the interaction distance $r$. For fast-decaying coupling terms, say e.g. $ j_r, \Delta_r \sim \mathrm{e}^{-|r|} $, the nearest-neighbor physics is largely recovered. However, the system with algebraic decay $ j_r \sim |r|^{-\alpha} \,, \ \Delta_r \sim |r|^{-\beta} $ (where $\alpha, \beta > 1$ throughout this paper) represents an example of infinite-order model \cite{Fu2021} and displays novel and interesting phenomena \cite{Viyuela2015,Lepori2016,Vodola2014,AD17,Fraxanet21,Mahyaeh18,Giuliano18}. Long-range extensions are not just theoretically enticing, but experimentally motivated. Superconductor-semiconductor (SC-SM) nanowires actually present next-to-nearest neighbor (and beyond) corrections to the standard NN hopping and pairing terms \cite{Pan2021}, and LR Kitaev models have moreover been proposed as effective descriptions of periodically driven (Floquet) systems \cite{Benito14,Li17}. On top of that, the LR Kitaev chain approximates the long-range Ising model \cite{Jaschke17}, experimentally implementable on currently available quantum simulation platforms \cite{Periwal21}.

%In this paper, we are going to make a substantial step towards the understanding of the interplay between long-range couplings and topology by characterizing the relation between (zero-energy) edge states in the Kitaev chain and the value of the bulk topological index $w$ at the quantum critical point. By this study, we will show that long-range couplings may induce anomalous value of the bulk topological index at criticality ($w_{c}$) with respect to the nearest neighbor case.

%The purpose of this letter is twofold. On the one hand, a novel analytical method is used to establish existence and decay of edge states, including Majoranas, both in the NN and LR setting. On the %other hand, the bulk index $w$ is assigned a meaning at the quantum critical point (where it is usually ill-defined). The latter can then sometimes be ascribed to either of the two quantum phases %(trivial vs topological). In such cases, one speaks of an \textit{asymmetric quantum critical point}.

Independently of the interaction range, finite-size (or semi-infinite) topological superconductors usually exhibit compactly supported modes about their edges. Much of the boundary physics is related to such \textit{edge states} and their decay in the bulk. Their analytical or numerical exploration is therefore a problem of interest, and even more so in the KC, where they are the sought-after Majoranas. Various detection methods are available in the literature: exact diagonalization, finite difference equations, transfer matrices \cite{Suraj16,DeGottardi13} and so on \cite{Jaeger20,AD17}. Yet, most of these methods are numerical and can only target a finite chain. The study of universal scaling behaviour, however, requires to explicitly address systems in the thermodynamic limit. We thus introduce a different technique, \emph{the scattering approach}, capable of describing topological states in the (semi)-infinite problem. The scattering approach proposes to use a linear combination of adequately modified bulk-eigenstates, i.e., the known solutions of the eigenvector problem in the thermodynamic limit, to construct the desired (bound) edge-states \cite{GP2013,ScattBraunlich,ReedSimonIII}. This paradigm allows us to analytically study Majorana zero-modes (MZMs) both in the NN and long-range settings, recovering the expected exponential decay in the first case, and showing \textit{softening} in the second one.

By means of the scattering approach, we also substantially deepen our understanding of the interplay between long-range couplings and topology by establishing a relation between (zero-energy) edge states in the Kitaev chain and the value of the bulk topological index $w$ at the quantum critical point. The index $w$, originally defined for the NN chain, can be extended to the long-range (LR) case\,\cite{AD17}. It attains an integer non-zero (zero) value in the topological (trivial) phase independently of the interaction range. The quantum phase transition is achieved by tuning a chemical potential $ \mu $. At the transition point, i.e., when $\mu$ attains a critical value $\mu_c$, $w$ is formally ill-defined. This can be remedied via a straightforward redefinition, as we show. The newly introduced critical value $w_c$ of $ w $ is $w_{c}=1/2$ in the NN model, perfectly interpolating between the trivial ($w=0$) and topological ($w=1$) phase, see Appendix \ref{app:windCrit}. 

The same is not generally true in the LR picture, where appropriately chosen decay exponents $\alpha, \beta$ can produce $w_{c}=0$. This formal result hints at the possibility of assigning the critical point to one of the two phases, rather than leaving it out of the classification as usual. We thus identify the values of $ \alpha, \beta$ yielding $ w_c = 0 $ and propose an edge interpretation of the phenomenon. This interpretation only applies to the \textit{hopping-dominated regime}, where $ \alpha < \beta $ and $\alpha$ sufficiently small.

\section{Bulk model}

The anisotropic long-range Kitaev chain consists in a 1D array of $ N $ sites hosting spinless fermions. The $i$th site fermionic operators are $ c_i, c_i^{\dagger} $, satisfying the usual canonical anticommutation relations. Its Hamiltonian reads
\begin{equation}
	\label{eq:ham}
	H= - \mu \sum_{i}(1-2c^{\dagger}_{i}c_{i}) -\sum_{i,r}(j_{r}c^{\dagger}_{i}c_{i+r}+\Delta_{r}c^{\dagger}_{i}c^{\dagger}_{i+r}+h.c.) \,,
\end{equation}
where $r$ represents the interaction distance, and 
\begin{equation}
		\label{eq:CouplingPairing}
	j_r \coloneqq j r^{-\alpha} \,, \qquad \Delta_r \coloneqq \Delta r^{- \beta} \,,
\end{equation}
with $ \alpha, \beta > 1 $ and $ j = \Delta = 1 $ in the following. Notice that $ r \in \{1,2,..., \infty \} $ in the infinite chain case, whereas one usually assumes $ r < N/2 $ for finite sample-size.

The bulk problem is solved exactly by successive application of a Fourier and Bogoliubov transform. We define the former as
\begin{align}
	\label{eq:fTrans}
	c_{r}=\frac{\eul^{i\frac{\pi}{4}}}{\sqrt{N}}\sum_{n=-\frac{N}{2}}^{\frac{N}{2}}c_{q_n}e^{iq_{n}r} \,, \qquad q_n = \frac{2 \pi n}{N} \,,
\end{align}
where the extra phase prevents the appearance of imaginary units in the transformed Hamiltonian. The latter then reads
\begin{equation}
	\label{eq:hamF}
	H=-2\sum_{k}(c^{\dagger}_{k}c_{k}
	-c_{-k}c^{\dagger}_{-k})\varepsilon_{k}+(c^{\dagger}_{k}c^{\dagger}_{-k}+c_{-k}c_{k})\Delta_{k} \,,
\end{equation}
where $ \varepsilon_k \coloneqq \mu - j_k $ and
\begin{align}
	\label{eq:epsDelta}
	j_{k} &= \sum_{r=1}^{\infty} \cos(kr) r^{-\alpha} = Cl_\alpha (k) \,, \nonumber \\
	\Delta_{k} &= \sum_{r=1}^{\infty} \sin(kr) r^{-\beta} = S_\beta (k) \,, 
\end{align}
where $Cl_\alpha (k)$ and $S_\beta (k)$ are Clausen functions of the first and second kind of index $\alpha, \beta$. The final diagonal form
\begin{equation}
	\label{eq:hamDiag}
	H= \sum_{k} \omega_{k}\left(\gamma^{\dagger}_{k}\gamma_{k}-\frac{1}{2}\right)
\end{equation}
with eigenvalues (bands)
\begin{equation}
	\label{eq:omega}
	\pm \omega_{k}= \pm \sqrt{\varepsilon_{k}^{2}+\Delta_{k}^{2}}
\end{equation}
is then achieved via the Bogoliubov transformation
\begin{equation}
	\label{eq:bgTrans}
	c_{k} = u_{k} \gamma_{k} - v^{*}_{-k}\gamma^{\dagger}_{-k} \,,
\end{equation}
where
\begin{equation}
	\label{eq:bgFactors}
	(u_{k},v_{k})=\left(\cos\frac{\theta_{k}}{2}, \sin\frac{\theta_{k}}{2}\right) \,
\end{equation}
and $ \theta_k $ is known as the Bogoliubov angle
\begin{equation}
	\label{eq:bgAngle}
	\tan\theta_{k} = \frac{\Delta_{k}}{\varepsilon_{k}} \,.
\end{equation} 

Equation \eqref{eq:hamF} can be recast into Bogoliubov-de Gennes form
%\begin{align}
%	\label{eq:hamFBog}
	%H &= -2 \sum_{k}
	%\begin{pmatrix}
	%	c^{\dagger}_k & c_{-k} 
	%\end{pmatrix}
	%\begin{pmatrix}
	%	\varepsilon_k & \Delta_k \\
	%	\Delta_k & - \varepsilon_k
	%\end{pmatrix}
	%\begin{pmatrix}
	%	c_k \\
	%	c^\dagger_{-k}
	%\end{pmatrix} H &=\nonumber \\
	%&\equiv -2 \sum_{k} \vec{c}^{\, \dagger}_k H(k) \vec{c}_k \,,
%\end{align}
$H \equiv -2 \sum_{k} \vec{c}^{\, \dagger}_k H(k) \vec{c}_k \,,
$ where $ \vec{c}_k = (c_k \ c^{\dagger}_{-k})^T $ and 
\begin{equation}
	\label{eq:hamk}
	H(k) = \vec{h} (k) \cdot \vec{\sigma} \,,
\end{equation}
with $ \vec{\sigma} = (\sigma_x, \sigma_y, \sigma_z) $ the vector of Pauli matrices and $ \vec{h} (k) = ( \Delta_k, \ 0, \ \varepsilon_{k} ) $. Equation \eqref{eq:hamk} grants particle-hole symmetry of $H$ and allows for a handy definition of the bulk index
\begin{equation}
	\label{eq:bulkInd}
	w = - \frac{1}{2 \pi} \oint \mathrm{d} \theta_k = \frac{1}{2 \pi} \int_{-\pi}^\pi \mathrm{d} k \frac{\partial_k \hat{h}_z (k)}{\hat{h}_x (k)} \,,
\end{equation}
where $ \hat{h} (k) \coloneqq \vec{h} (k) / \Vert \vec{h} (k) \Vert $. The rightmost member of Eq.\,\eqref{eq:bulkInd} is the winding number of the curve $ \vec{h}(k) $ about the origin.

\begin{figure*}
	\subfigure[$ \mu >\mu_{c}$]{\label{fig:1a}\includegraphics[width=.32\linewidth]{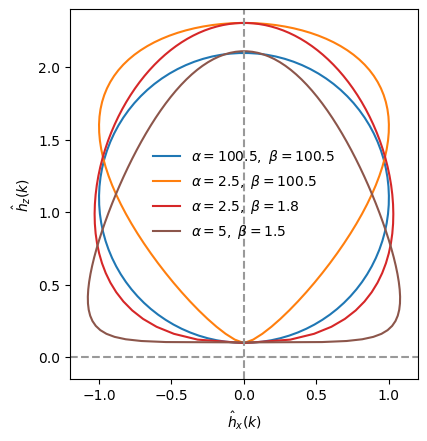}}
	%\hspace{1cm}
	\subfigure[$ \mu = \mu_c $]{\label{fig:1b}\includegraphics[width=.32\linewidth]{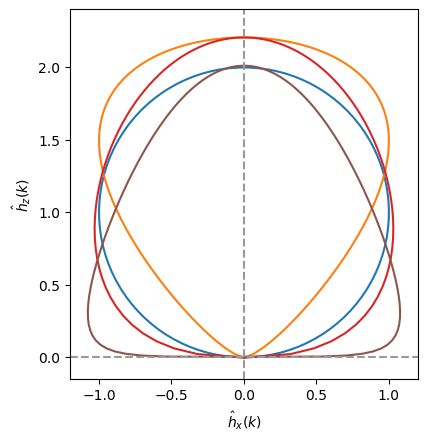}}
	\subfigure[$ \mu < \mu_c $]{\label{fig:1c}\includegraphics[width=.32\linewidth]{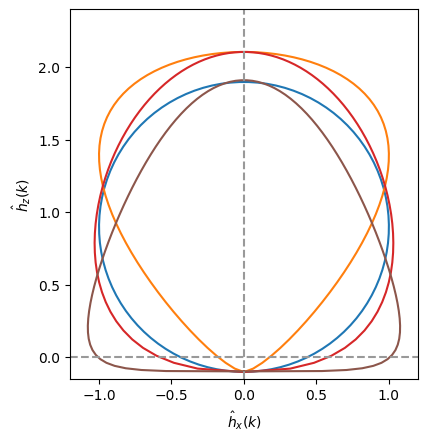}}
	\caption{\textbf{Visualization of the bulk index.} Depicted is the curve $ \vec{h} (k) \,, \ k \in \{ 0, 2 \pi \} $, whose winding about the origin constitutes the bulk index $w$, for various values of $  \alpha, \beta$. The left (right) panel represents the situation in the trivial (topological) phase. At criticality, central panel, curves intersect the origin and the winding number is formally ill-defined. \label{Fig1}}
\end{figure*}

As anticipated, $ w=0 $ in the trivial phase, where a lack of Majorana zero-modes at the edges is expected. By contrast, $w>0$ and integer in the topological phase. In principle, $ w $ is undefined at criticality, since the curve $ \vec{h}(k) $ intersects the origin. Yet, a reasonable definition of $ w_c $ is obtained by simply reading the integrals in Eq.\,\eqref{eq:bulkInd} as principal values. In the rest of the article, this point of view will be adopted to compute $ w_c $ and argue when $ w_c = 0,\,1/2$ or $1$. A drawing of $ \vec{h}(k) $ winding about the origin in the trivial, critical and topological regimes is reported in Fig.\,\ref{Fig1}

\section{Edge model and scattering approach}

An edge is introduced by cutting out the left-side of the chain, namely restricting position space from $ \mathbb{Z} $ to $ \mathbb{N} $. Majoranas are hence only found at the \enquote{left end} of the chain. Bulk eigenstates have already been identified as
%\begin{align}
%	\label{eq:bulkEigen}
%	\gamma_k &= \cos(\theta_k / 2) c_k + \sin(\theta_{-k} / 2) c^\dagger_{-k} \nonumber \\ 
%	&= \cos(\theta_k / 2) c_k - \sin(\theta_{k} / 2) c^\dagger_{-k} \,,
%\end{align}
the $ \gamma $ operators of Eq.\,\eqref{eq:bgTrans}. These operators satisfy the eigenvalue equation $ [H,\gamma_k] = - \omega(k) \gamma_k $. %rather than $ H \gamma_k = - \omega(k) \gamma_k $, and we refer to them as \textit{states}, in the second-quantized fashion. 
Edge modes of energy $ E $ are states $\hat{\psi}$ supported on the right half-line satisfying $ [\hat{H}, \hat{\psi}] = E \hat{\psi} $, with $ \hat{H} $ restriction of the bulk operator to the new position space. A Majorana is then a \textit{zero-energy edge mode} $\hat{M}$, solution of the equation
\begin{equation}
	\label{eq:MajoDef}
	[\hat{H}, \hat{M}] = 0 \cdot \hat{M} = 0 \,,
\end{equation}
i.e., an edge state commuting with the restricted Hamiltonian.

Various methods to solve Eq.\,\eqref{eq:MajoDef} exist. The conceptually simplest one consists in writing $H$ in its position-space BdG form
%\begin{equation}
%	\label{eq:posSpaceBdG}
%	H = 
%	\begin{pmatrix}
%		c^\dagger_{1} & \ldots & %c^\dagger_{N} & c_1 & \ldots & c_N 
%	\end{pmatrix}
%	H_{BdG}
%	\begin{pmatrix}
%		c_1 \\
%		\vdots \\
%		c_N \\
%		c^\dagger_1 \\
%		\vdots \\
%		c^\dagger_N
%	\end{pmatrix} \,
%\end{equation}
and finding the zero-eigenvalue eigenvectors of the $ 2N \times 2N $ matrix $ H_{\rm BdG} $. This method is feasible as long as $ H_{\rm BdG} $ is a banded Toeplitz matrix. Such a feature is lost when couplings become all-to-all, as for the power-law decaying interactions treated here. Then, $ H_{\rm BdG} $ is a full matrix and brute-force numerics fail for relatively small chain-size $N$. This hinders the realization of an effective finite-size scaling capable to describe the behaviour of the half-infinite system. Analytical solutions of Eq.\,\eqref{eq:MajoDef}, however, have difficulties going beyond NNN interactions, e.g., Ref.\,\cite{Mahyaeh18} with Lieb-Schulz-Matthis method, or do so at the price of engaging in very heavy computations \cite{Jaeger20,Jones2022}.

By contrast, we wish to propose a more straigthforward technique, known in the mathematical physics literature as the \enquote{scattering approach}\,\cite{GP2013,ScattBraunlich,ReedSimonIII}, that requires nothing but the solution of the bulk model (given above) and few reasonable calculations. Most noticeably, the NN and LR results presented below will be obtained directly in the thermodynamic limit. This desirable feature is rare in the literature, but not unheard of: Refs.\,\cite{Patrick17,Jones2022} achieve it using methods similar to ours.  

Let us illustrate here the philosophy behind this approach. The fully general procedure is only reported in App.\,\ref{app:GenScattApp}, but the central idea follows. Eigenstates $ \gamma^{\star}_{\pm k} $ ($ \star $ standing for \enquote{nothing} or $ \dagger $) of the translation-invariant bulk Hamiltonian $ H $ are, in essence, plane waves. Consider now $ \gamma^\star_\kappa $, where $ k \in \mathbb{R} $ was substituted by $ \kappa \in \mathbb{C} $. These are still formal solutions of $ [H,\gamma^\star_\kappa] = \pm \omega (\kappa) \gamma^\star_\kappa $, yet they cannot be considered proper eigenstates: akin to growing or decaying exponentials, they diverge towards one of the two \enquote{ends} of the infinite chain. However, when retaining only the right half of the line, \enquote{evanescent} modes (those with $ \mathrm{Im} \ \kappa > 0 $) become physical, and localize around the boundary, like the sought after edge (bound) states. This reasoning prompts, for an edge mode of positive energy $E$, the ansatz
\begin{equation}
	\label{eq:ScattAnsatz}
	\psi_s (E) = \sum_{j \in J} A_j \hat{\gamma}_{\kappa_j}^{\dagger} + B_j \hat{\gamma}^\dagger_{-\kappa_j} \,,
\end{equation} 
where $J$ is the (possibly empty) set of labels $j$ of complex momenta $ \kappa_j $ such that $ \omega (\kappa_j) = E $ and $ \mathrm{Im} \ \kappa_j > 0 $ (evanescent wave). Modes $ \gamma_\kappa, \ \gamma_{-\kappa} $ do not enter in Eq.\,\eqref{eq:ScattAnsatz} by assumption of positive energy. By contrast, the operators appearing carry a hat to signify restriction to the half-line.

Majoranas are thus obtained from the ansatz above by setting $E=0$. In this case, $ \gamma^\dagger_{-\kappa} $ comes to coincide with its particle-hole conjugate $ \gamma_\kappa $, and one can thus tweak Eq.\,\eqref{eq:ScattAnsatz} to
\begin{equation}
	\label{eq:ScattAnsatzMajo}
	\psi_s (0) \equiv \hat{M} = \sum_{j \in J} A_j \hat{\gamma}_{\kappa_j}^{\dagger} + B_j \hat{\gamma}_{\kappa_j} \,.
\end{equation} 

A few remarks. First, $ \hat{\gamma}^\dagger_\kappa $ is not an eigenstate of momentum when $ \kappa \in \mathbb{C}$. Unable to write it in momentum space, we just consider the original $ \gamma^\dagger_k $ ($k$ real) in position space and produce the \enquote{wave function} of $ \gamma^\dagger_\kappa $ by $ k \mapsto \kappa $ substitution and restriction to $ \mathbb{N} $
\begin{align}
	\label{eq:GammaKappa}
	\hat{\gamma}^\dagger_\kappa &= C \left[ - \eul^{- \im \frac{\pi}{4}} \sin \left( \frac{\theta_{\kappa}}{2} \right) \sum_{s= 0}^{+ \infty} c_s \eul^{\mathrm{i} \kappa s} \right. \nonumber \\ 
	&+ \left. \eul^{\im \frac{\pi}{4}} \cos \left( \frac{\theta_\kappa}{2} \right) \sum_{s= 0}^{+ \infty} c^\dagger_s \eul^{\mathrm{i} \kappa s} \right] \,,
\end{align}
with $C \in \mathbb{R}$ a normalization constant.

Second, particle-hole conjugation, here denoted $ \mathcal{P} (\cdot) \mathcal{P}^{-1} $, acts like $ (\cdot)^\dagger $ on linear combinations of $ c_l, c_l^\dagger $. It is customary to deem a state $\psi$ particle-hole symmetric if $ \psi^\dagger = \psi $. This relation cannot be satisfied by our bulk modes, due to the extra phase introduced in the Fourier transform. In the following, we \enquote{rotate} such modes back to the standard convention, i.e., work with
\begin{align}
	\chi_+ (\kappa) \coloneqq e^{- \im \pi/4} \gamma^\dagger_\kappa \,, &\qquad \chi_- (\kappa) \coloneqq \eul^{\im \pi/4} \gamma_\kappa \,, \nonumber \\ 
	\varphi_- (\kappa) \coloneqq \eul^{\im \pi/4} \gamma_{-\kappa} \,, &\qquad \varphi_+ (\kappa) \coloneqq \eul^{- \im \pi/4} \gamma_{-\kappa}^\dagger \label{eq:ChiPhiDef} \,.
\end{align}
Finally, the attentive reader may argue that edge eigenstates akin to Eq.\,\eqref{eq:GammaKappa} could have been obtained by substituting the discrete Fourier transform in Eq.\,\eqref{eq:fTrans} with a Laplace transform $ c_r \mapsto c_\kappa \,, \kappa \in \mathbb{C} $. The \textit{scattering approach} is equivalent to solving the edge problem anew via Laplace transform, but here edge solutions are simply obtained from their bulk (Fourier) counterpart by the extension $ k \in \mathbb{R} $ to $ \kappa \in \mathbb{C} $.

\section{Edge states of the NN model}

Let us apply the outlined procedure to the original Kitaev chain. Start by replacing $ \hat{\gamma}^\dagger, \hat{\gamma} $ with $ \hat{\chi}_+, \hat{\chi}_- $ in Eq.\,\eqref{eq:ScattAnsatzMajo}, to recover the standard particle-hole symmetry as explained above. This results in
\begin{equation}
	\label{eq:ScattAnsatzMajoNN}
	\hat{M} = \sum_{j \in J} A_j \hat{\chi}_+ (\kappa_j) + B_j \hat{\chi}_- (\kappa_j) \,.
\end{equation} 
Let us focus on $ \chi_+ $ first. By its definition and Eq.\,\eqref{eq:GammaKappa}, it explicitly reads
\begin{align}
	\label{eq:chiPlus}
	\chi_+ (\kappa) &= C \left[ \mathrm{i} \sin \left( \frac{\theta_\kappa}{2} \right) \sum_{s= - \infty}^{+ \infty} c_s \eul^{\mathrm{i} \kappa s} \right. \nonumber \\
	&+ \left. \cos \left( \frac{\theta_\kappa}{2} \right) \sum_{s= - \infty}^{+ \infty} c^\dagger_s \eul^{\mathrm{i} \kappa s} \right] \,.
\end{align}
The first step towards determining $ \hat{M} $ is solving $ \omega(\kappa) = 0 $. By the definitions in Eq.\,\eqref{eq:epsDelta} one has $ \varepsilon(k) = \mu - \cos(k) $ and $ \Delta_k = \sin(k) $ in the NN case, so that
\begin{equation}
	\omega (\kappa) = \sqrt{\mu^2 + 1 - 2 \mu \cos (\kappa)} \,,
\end{equation} 
see Eq.\,\eqref{eq:omega}. Then $ \omega(\kappa) = 0 $ only if
\begin{equation}
	\label{eq:kappas}
	\begin{array}{lcll}
		\hat{\kappa}_1 = & \im \ \rm{arccosh} \left( \frac{\mu^2 + 1}{2 \mu} \right) & \,, & (\mu >0) \\
		\hat{\kappa}_2 = & \pi + \im \ \rm{arccosh} \left( \frac{\mu^2 + 1}{-2 \mu} \right) & \,, & (\mu < 0) \,,
	\end{array}
\end{equation}
where each solution is actually double ($ \rm{arccosh} $ is two-valued).

Let us consider $ \mu > 0 $ for simplicity; $\mu < 0 $ is analogous in all respects. $ \chi_+ (\hat{\kappa}_1) $ can only represent a Majorana if it is particle-hole symmetric. Eq.\,\eqref{eq:chiPlus} entails $ \mathcal{P} \chi_+ (\hat{\kappa}_1) \phs^{-1} = \chi_+ (\hat{\kappa}_1) $ if and only if
\begin{equation}
	\cos (\theta_{\hat{\kappa}_1} / 2 ) = - \im \overline{\sin( \theta_{\hat{\kappa}_1} / 2 )} \,,
\end{equation}
condition that is met when $ \theta_{\hat{\kappa}_1} \to - \im \infty $. Recalling Eq.\,\eqref{eq:bgAngle}, it must then be $ \Delta_{\hat{\kappa}_1} / \varepsilon_{\hat{\kappa}_1} = - \im $ or equivalently
\begin{equation}
	\label{eq:ImposingPHS}
	\varepsilon_{\hat{\kappa}_1} - \im \Delta_{\hat{\kappa}_1} = 0 \,.
\end{equation}
However, by direct substitution of Eq.\,\eqref{eq:kappas} into $ \Delta_k = \sin(k) \,, \ \varepsilon(k) = \mu - \cos(k) $,
\begin{equation}
	\frac{\Delta_{\hat{\kappa}_1}}{\varepsilon_{\hat{\kappa}_1}} = \im \, \rm{sgn} (\mu^2 - 1) \,,
\end{equation}
implying that Majoranas exist for $ 0 < \mu < \mu_c = 1 $. Repeating the computations for $ \mu < 0 $, using $ \hat{\kappa}_2 $ rather than $\hat{\kappa}_1$, yields existence for $ -1 < \mu < 0 $. We have thus identified the topological phase as $ -1 < \mu < 1 $, in agreement with the expectation obtained by the bulk topological index; see Fig.\,\ref{Fig1}.

A simple substitution of $ \hat{\kappa}_1 $ into $ \chi_+ $ now yields
\begin{align}
	\label{eq:MajoChi}
	\chi_+^{(1,2)} (\hat{\kappa}_1) &= C (\hat{\kappa}_1) \sum_s (c_s + c_s^\dagger) \eul^{ \im \hat{\kappa}_1 s} \nonumber \\
	&=  C (\hat{\kappa}_1) \sum_s (c_s + c_s^\dagger) \mu^{\pm s} \,, \ C(\hat{\kappa}_1) \in \mathbb{R} \,,
\end{align}
where the normalization constant diverges. Here, this fact carries no physical consequence, but we will have to bear it in mind for our long-range analysis. The two solutions $ \chi_+^{(1,2)} $ correspond to the two branches of $ \rm{arccosh} $.

The entire procedure can be repeated for $ \chi_- (\kappa)$, obtaining ($0<\mu<1$)
\begin{equation}
	\label{eq:ChiMajo}
	\chi_-^{(1,2)} (\hat{\kappa}_1) = \im D (\hat{\kappa}_1) \sum_s (c_s - c_s^\dagger) \mu^{\mp s} \,, \ D(\hat{\kappa}_1) \in \mathbb{R} \,.
\end{equation}
Having determined all of the linearly independent zero-energy solutions for positive $\mu$, we combine them as
\begin{equation}
	M = A_1 \chi_+^{(1)} + A_2 \chi_+^{(2)} + B_1 \chi_-^{(1)} + B_2 \chi_-^{(2)} \,.
\end{equation}
The states $ \chi_\pm^{(1,2)} (\hat{\kappa_1}) $ are still defined on the entire 1D lattice, and restriction to $ \mathbb{N} $ is only possible upon discarding the divergent waves $ \chi_+^{(2)} , \chi_-^{(1)} $. The final form of the Majorana edge modes, see Eq.\,\eqref{eq:ScattAnsatzMajoNN}, is hence
\begin{equation}
	\label{eq:finalMajo}
	\hat{M} = A_1 \sum_{s=0}^{\infty} (c_s + c_s^\dagger) \mu^s + \im B_2 \sum_{s=1}^{\infty} (c_s - c_s^\dagger) \mu^s \,,  
\end{equation}
where $A_1,B_2 \in \mathbb{R}$ have absorbed the normalization constants $ C(\hat{\kappa}_1), D(\hat{\kappa}_1) $, cf.\,Eqs.\,\eqref{eq:MajoChi} and \eqref{eq:ChiMajo}. Albeit derived with completely different methods, the last equation is in perfect agreement with Ref.\,\cite{KitaevChain}.

In closing, let us comment on how many independent Majorana modes exist according to Eq.\,\eqref{eq:finalMajo}. There are two complex coefficients $ A_1 $ and $B_2$. When interactions are nearest neighbor, the boundary condition consists in a single equation, which fixes one of them. The remaining coefficient is determined by normalization. One must thus conclude that no more than one Majorana mode can localize at the left edge of the semi-infinite chain, in agreement with all existing literature.

\section{The long-range case, algebraic decay of Majoranas}

There is no fundamental obstruction to extending the methods above to the long-range case. However, identifying edge states of energy $E$ requires solving
\begin{equation}
	\label{eq:generalKappa}
	\omega (\kappa) = \sqrt{(\mu - Cl_{\alpha} (\kappa))^2 + S_\beta^2 (\kappa)} = E \,,
\end{equation}
far from an easy task. Nonetheless, the expertise matured in solving the NN problem and some carefully chosen approximations will allow us to deduce the qualitative behaviour of Majorana edge modes, and in particular their algebraic decay \cite{Jaeger20, Vodola2014, AD17, Vodola2015} close to a quantum critical point.

The prescriptions of the scattering approach and particle-hole symmetry of the zero-modes impose the following form for a Majorana state
\begin{equation}
	\label{eq:LRMajo}
	\hat{M} = \sum_{j \in J} \sum_{s=0}^{\infty} \cos \left( \frac{\theta_{\kappa_j}}{2} \right) \left[ A_j (c_s + c_s^\dagger) + \im B_j (c_s - c_s^\dagger) \right] \eul^{\im \kappa_j s} \,,  
\end{equation}
where $ A_j, B_j \in \mathbb{R} $ and $ \omega (\kappa_j) = 0, \ \forall j \in J $. The set $\{\kappa_{j}\}$ contains all possible solutions to Eq.\,\eqref{eq:generalKappa} with $E=0$. The exact $ \kappa_j $ are not known,  due to the difficulty in solving Eq.\,\eqref{eq:generalKappa}, and $ \cos ( \theta_{\kappa_j} / 2 ) $ diverges for zero-energy solutions, as seen in the NN case.

In analogy with the NN case, we retain only the \enquote{even} half of Eq.\,\eqref{eq:LRMajo}, and pick the boundary condition in such a way that $ A_j = A_k, \ \forall j,k $, leading to
\begin{equation}
	\label{eq:LRMajoReduced}
	\hat{M} = \sum_{j \in J} \sum_{s=0}^{\infty} \cos \left( \frac{\theta_{\kappa_j}}{2} \right) (c_s + c_s^\dagger) \eul^{\im \kappa_j s} \,.  
\end{equation}
Since we are interested in investigating the topological properties at the critical point, which will be related to the universal scaling of the topological states, we introduce a low-energy approximation by expanding the coupling coefficients at small $\kappa$
\begin{align}\label{eq:EpsDeltaExpansionGeneric}
	\varepsilon_{\kappa} &= \tau - \varepsilon_\alpha \kappa^{\alpha -1} + \varepsilon_2 \kappa^2 + \mathcal{O} (\kappa^4) \,, \nonumber \\
	\Delta_{\kappa} &= \delta_\beta \kappa^{\beta - 1} + \delta_1 \kappa + \mathcal{O} (\kappa^3) \,,
\end{align}
where $ \tau= \mu - \mu_c $ and $ \varepsilon_{(\cdot)}, \delta_{(\cdot)} $ are currently unspecified complex coefficients. Inserting the expansions in Eq.\,\eqref{eq:EpsDeltaExpansionGeneric} into Eq.\,\eqref{eq:generalKappa}, we will obtain an explicit expression for the \textit{inverse localization lengths} $ \kappa_j $. 

Before following the aforementioned  procedure two remarks are in order:
(i) Since the expansions in Eq.\,\eqref{eq:EpsDeltaExpansionGeneric} have a finite convergence radius\,\cite{L80,L81}, they reproduce the non-analytic (log-type) behaviour of the true coupling functions $ \varepsilon_\kappa, \Delta_\kappa $ in Eq.\,\eqref{eq:epsDelta}. Therefore, infinitely many $\kappa_{j}$ solutions emerge, which is a known consequence of the infinite coordination number of long-range interactions\,\cite{AD17}. (ii) Also in the long-range case, the expression for $ \cos ( \theta_{\kappa_j} / 2 ) $ diverges as the limit $E\to 0$ is approached. Extra care will be demanded to treat this divergence.  %This result for $\kappa_j$ can then be inserted in Eq.\,\eqref{eq:LRMajoReduced}. Extra care is demanded by the divergence of $ \cos ( \theta_{\kappa_j} / 2 ) $ as $ E \to 0 $. We hereby notice that, by Eq.\,\eqref{eq:EpsDeltaExpansionGeneric} and the definition of $ \omega (\kappa) $, small $ \kappa $ solutions are expected to be the only relevant ones close to criticality ($ \tau \to 0 $).

Our approach starts by inserting Eq.\,\eqref{eq:EpsDeltaExpansionGeneric} into Eq.\,\eqref{eq:generalKappa}, which yields
\begin{align}
	\label{eq:OmegaLambda}
	\omega^2 &= \tau^2 - 2 \tau \varepsilon_\alpha \kappa^{\alpha -1} + \varepsilon_\alpha^2 \kappa^{2(\alpha -1)} \nonumber \\
	&+ \delta_\beta^2 \kappa^{2(\beta - 1)} + (\delta_1^2 + 2 \tau \varepsilon_2) \kappa^2 + ... = \lambda^2 \,,
\end{align}
having set $ E = \lambda^2 $, with $ \lambda $ a \textit{small yet finite} constant. Keeping $ \lambda \neq 0 $ prevents us from hitting certain infinities, thus making the computations technically manageable. 

Three different regimes can be identified, depending on the leading power of $ \kappa $ in Eq.\,\eqref{eq:OmegaLambda}
\begin{enumerate}
	\item \textit{Almost finite-range:} if $ \alpha > 3 $ and $ \beta > 2 $, then $ \kappa^2 $ leads. The universal scaling of quantities in this regime is akin to the nearest-neighbor case.
	
	\item \textit{Hopping-dominated:} if $ \alpha < 3 $ and $ \alpha < \beta $, then $ \kappa^{\alpha -1} $ (embodying the \textit{hopping} term of the Hamiltonian) leads.
	
	\item \textit{Pairing-dominated:} if $ \beta < 2 $ and $ \alpha > \beta $, then $ \kappa^{2(\beta -1)} $ (embodying the \textit{pairing} term of the Hamiltonian) leads.
\end{enumerate}
In the \textit{almost finite-range} case, there is only one solution $ \kappa_1 $ to the leading order of Eq.\,\eqref{eq:OmegaLambda}, and everything reduces to the NN case. By contrast, in the other two cases, one has as many $ \kappa_j $ as there are roots to $ \kappa \simeq \tau^{1/(\alpha - 1)} $ or $ \kappa \simeq \tau^{1/(\beta - 1)} $. For $ \alpha, \beta $ irrational (which is the general case), this number is infinite and the solutions lie homogeneously on a circle $ \mathcal{C}_\rho $ of fixed radius $ \rho $ in the $ \kappa $-complex plane. 

The Majoranas can now be constructed, upon evaluation of $ \cos (\theta_{\kappa} / 2) $ at $ \kappa_j $. Performing the operation with great care yields, in each of the three regimes above
\begin{equation}
	\label{eq:CosThetaK}
	\cos \left( \frac{\theta_{\kappa_j}}{2} \right) \propto 
	\begin{cases}
		\kappa_j \quad &\mathrm{if}\, \alpha > 3 \ \text{and} \ \beta > 2 \\
		\kappa_j^{\alpha-1} \quad &\mathrm{if}\, \alpha < 3 \ \text{and} \ \alpha < \beta \\
		\kappa_j^{\beta-1} \quad &\mathrm{if}\, \beta < 2 \ \text{and} \ \alpha > \beta
	\end{cases}
\end{equation}
up to the expected $\lambda$-dependent divergent prefactor. See Appendix \ref{app:LRcase} for details.

Let us focus specifically on the \textit{hopping-dominated} regime. Out of the infinitely many $ \kappa_j $, we retain only the physical solutions which decay on the correct half-space and plug Eq.\,\eqref{eq:CosThetaK} into Eq.\,\eqref{eq:LRMajoReduced}. Since the $\kappa_{j}$ solutions become dense on $\mathcal{C}_\rho$, we approximate the infinite sum in Eq.\,\eqref{eq:LRMajoReduced} by an integral, leading to (see Appendix \ref{app:LRcase} for details)
\begin{align}
	\label{eq:LRMajoIntegral}
	\hat{M} &= C \sum_{s=1}^{\infty} \int_{\mathcal{C}_\rho^+} \kappa^{\alpha - 1} (c_s + c_s^\dagger) \eul^{\im \kappa s} \mathrm{d} \kappa \\
	&\equiv C \sum_{s=1}^{\infty} f(s) (c_s + c_s^\dagger) \,,
\end{align}
where $ \mathcal{C}_\rho^+ $ represents the upper half of the radius-$\rho$ circle in the complex $\kappa$ plane and
\begin{equation}
	f(s) \coloneqq \int_{\mathcal{C}_\rho^+} \kappa^{\alpha - 1} \eul^{\im \kappa s} \mathrm{d} \kappa \,.
\end{equation}
Assuming $ s \to \infty $ (to study the physics deep in the bulk), by the saddle point method $ f(s) \propto s^{-\alpha} $. Hence
\begin{align}
	\label{eq:LRMajoFinal}
	\hat{M} &= \tilde{C} \sum_{s=1}^{\infty} s^{- \alpha} (c_s + c_s^\dagger) \,,
\end{align}
with $ \tilde{C} $ absorbing the $\lambda$-dependent divergence. Similar computations, reported in Appendix \ref{app:LRcase}, yield an $ s^{-\beta} $ decay in the \textit{pairing-dominated phase}, reproducing the findings of Ref.\,\cite{Jaeger20,Jones2022}, at least for $\alpha<3$ or $\beta<2$. A disagreement with said references emerges when $\alpha>3$ or $\beta>2$. This is a consequence of the different approximations employed in our treatment with respect to the one of Refs.\,\cite{Jaeger20,Jones2022}. This issue is further discussed in Appendix \ref{app:LRcase}.

\section{Universal scaling at criticality} \label{sec:BulkCrit}

After having determined the spatial decay of the Majorana edge states we can characterize their scaling at criticality and relate it to the values of $w_{c}$. Indeed, the critical bulk index $ w_c $ of the long-range model can vanish, signalling a discrepancy between the critical properties of the long-range model and the NN case, where $w_{c}=1/2$. The aim of this paragraph is to explain the genesis of this discrepancy, identify the values of $ (\alpha, \beta) $ s.t. $ w_c = 0 $ and propose an \enquote{edge} interpretation of the phenomenon.

\begin{figure}
	\frame{\includegraphics[width=.95\linewidth]{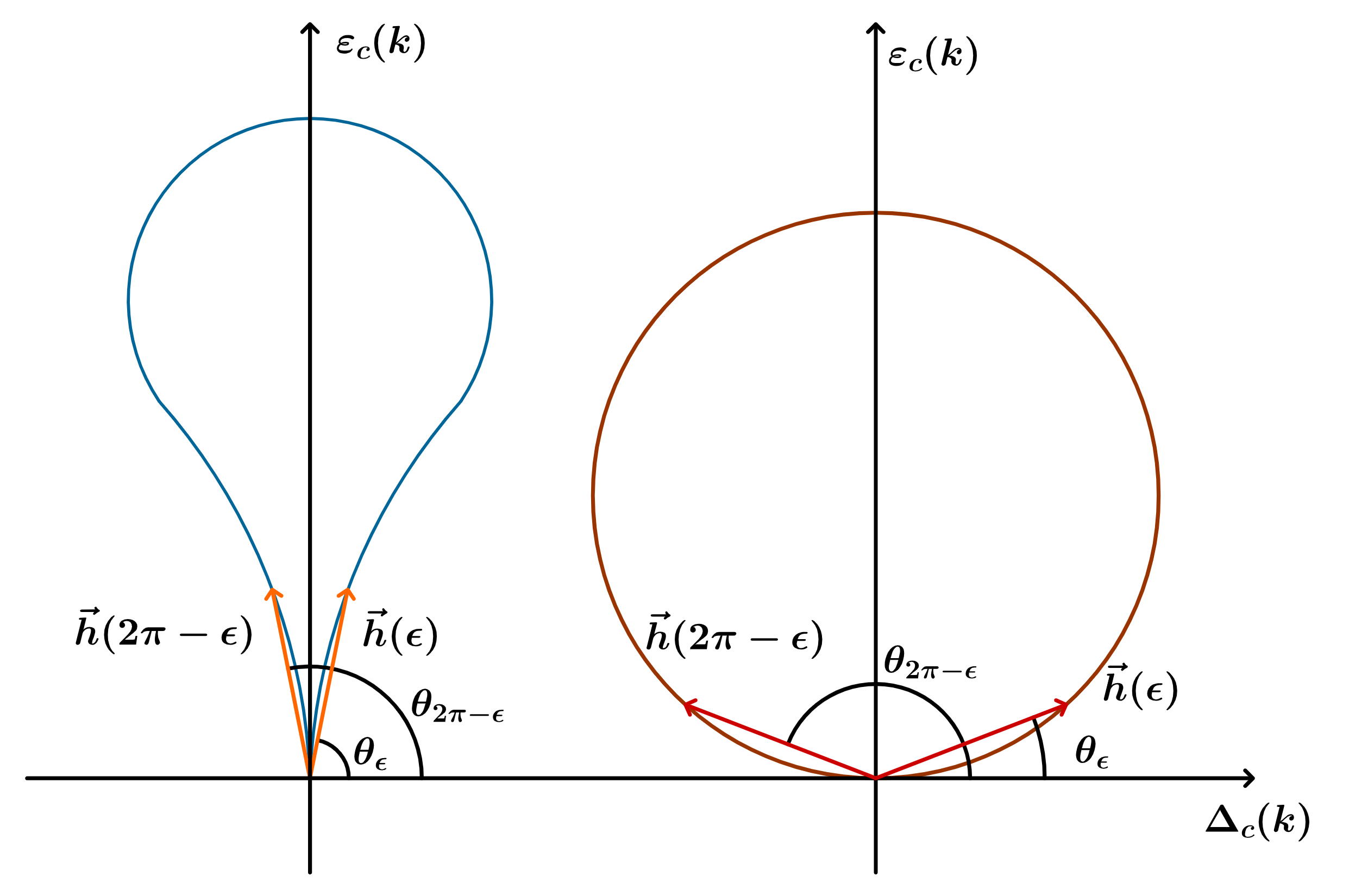}}
	\caption{\textbf{Computing $w_c$.} The left (right) panel represents the $ \varepsilon_k / \Delta_k \to +\infty \,, \ k \to 0 $ ($ \varepsilon_k / \Delta_k \to 0 \,, \ k \to 0 $) situation. In the first case, $ \vec{h} (\epsilon) $ and $ \vec{h} (2 \pi - \epsilon) $ become parallel as $ \epsilon \to 0 $. By contrast, they become anti-parallel in the opposite regime.}
	\label{fig2}
\end{figure}

Recall the definition \eqref{eq:bulkInd} of $w$, winding of the curve $ \vec{h} (k) = (h_x (k),0, h_z(k)) \,, \ k \in [-\pi, \pi] $ about the origin. By the fundamental theorem of calculus
\begin{equation}
	w = - \frac{1}{2 \pi} (\theta_{\pi} - \theta_{-\pi}) \,.
\end{equation}
This quantity is ill-defined when $ \mu = \mu_c $, because the curve intersects the origin, namely the point our angle is measured from. Nonetheless, we are free to redefine
\begin{equation}
	w \rvert_{\mu = \mu_c} \equiv w_c \coloneqq -\frac{1}{2 \pi} \lim_{\epsilon \to 0^+} (\theta_{\pi - \epsilon} - \theta_{- \pi + \epsilon} ) \,. \label{eq:CritIndex}
\end{equation}
Notice that a similar definition already appeared in \cite{Verresen18,Verresen20}. Simple observations allow $w_c$ as in Eq.\,\eqref{eq:CritIndex} to be computed \enquote{by naked eye}, see Fig.\,\ref{fig2}. The curve  $ (h_x (k), h_z(k)) = ( \Delta_k, \varepsilon_k ) $ is symmetric about the $z$ axis by $ \varepsilon_k = \varepsilon_{-k} $ and $ \Delta_k = - \Delta_{-k} $. Then, $ \theta_k = \pi - \theta_{-k} $ for all $ k > 0 $, and this holds true in the $ k \to \pi $ limit.

Now, allow us to take $ k \in [0, 2 \pi] $, so as to use the $ k \to 0 $ expansions \eqref{eq:EpsDeltaExpansionGeneric} of $ \varepsilon_k, \Delta_k $. Being at criticality, $ \tau = 0 $ and 
\begin{align}
	\label{eq:EpsDeltaExpansionCrit}
	\varepsilon_{k} &= \varepsilon_\alpha k^{\alpha -1} + \varepsilon_2 k^2 + \mathcal{O} (k^4) \,, \nonumber \\
	\Delta_{k} &= \delta_\beta k^{\beta - 1} + \delta_1 k + \mathcal{O} (k^3) \,,
\end{align}
both going to zero. If $ \Delta_{k} $ goes to zero faster (slower) than $ \varepsilon_{k} $, the vector connecting the origin to $ (h_x (k), h_z(k)) = ( \Delta_k, \varepsilon_k ) $ starts out vertical (horizontal), i.e., $ \theta_\epsilon = \pi/2 $ ($ \theta_{\epsilon} = 0 $). By the observations above, then $ \theta_{2 \pi - \epsilon} = \pi/2 $ ($ \theta_{2 \pi - \epsilon} = \pi $). In other words
\begin{equation}
	w_c = -\frac{1}{2 \pi} \lim_{\epsilon \to 0^+} (\theta_{2 \pi - \epsilon} - \theta_{\epsilon} ) =
	\begin{cases}
		0 \,, & (\varepsilon_\epsilon / \Delta_\epsilon) \to \infty \\
		\frac{1}{2} \,, & (\varepsilon_\epsilon / \Delta_\epsilon) \to 0 \,.
	\end{cases}
\end{equation}
In practice (see Appendix \ref{app:windCrit} for details), $ w_c = 0 $ is only achieved in the \textit{hopping-dominated} phase, for $ 1 < \alpha < 2 $. A similar discrepancy between the universal properties of the hopping- and pairing-dominated regimes was already noticed in the study of the Kibble-Zurek mechanism\,\cite{defenu2019universal}.

The \enquote{bulk} results above lend themselves to an \enquote{edge} interpretation. If $ \beta > \alpha $ and $ \alpha < 3 $, the inverse localisation length $ \kappa $, quantifying how bounded the Majorana modes are, is $ \kappa \propto \tau^{1 /(\alpha - 1)} $. Furthermore, if $ 1 < \alpha < 2 $ ($ 2 < \alpha < 3 $), then $ \kappa \propto \tau^ \gamma $ with $ \gamma > 1 $ ($ \gamma < 1 $). In the former (latter) case, bound states become \textit{{half-bound}} \cite{QuasiBound} faster (slower) than $\tau = \mu - \mu_c $ as $ \tau \to 0 $. Phrased differently, in the former (latter) situation bound states disappear \enquote{before} (as) criticality is reached. Deeming the transition point trivial when no Majoranas are present, this edge picture agrees with the critical bulk index $ w_c $ in assigning $ \mu = \mu_c $ to the trivial phase only if $ 1 < \alpha < 2 $.

\section{Conclusions and future directions}

To the best of our knowledge, this is the first time that the scattering approach was applied to the investigation of edge states in the Kitaev chain. This led to recovering the familiar Majorana zero-modes in the nearest-neighbor model, in perfect agreement with the celebrated results of Ref.\,\cite{KitaevChain}. The method has moreover proven itself flexible enough to treat fully connected power-law decaying couplings without additional difficulties. In the latter case, the emergence of non-analytic terms in the momentum-space couplings, see Eq.\,\eqref{eq:EpsDeltaExpansionGeneric}, generates infinitely many edge state solutions at low energy. The actual Majorana modes then emerge as a convolution of all the zero-energy solutions, leading to the well-known power-law decaying behaviour in real space. For $\alpha<3$ or $\beta<2$, such decay agrees with the one found in Ref.\,\cite{Jaeger20}.

Going beyond current literature, the scattering approach highlighted a novel form of universal scaling, displayed by the edge states as the critical point is approached. Indeed, the inverse localization lengths $\kappa$ vanish as a power-law $\kappa\propto \tau^{\gamma}$ when nearing criticality $\tau\to 0$. This mechanism, which entails that Majorana states become half-bound \cite{QuasiBound} at the critical point, is strongly influenced by the presence of long-range couplings. In particular, the scaling exponent $\gamma=1/(\alpha-1)$ of the hopping-dominated regime can grow very large, heavily smearing the (otherwise localised) edge modes. This softening of the Majorana states is simultaneously signalled by the bulk topological index remaining zero at criticality $w_{c}=0$, in contrast with the nearest neighbor case $w_{c}=1/2$.

Our findings represent a first step towards the characterization of the topological properties of long-range interacting systems, directly in the thermodynamic limit. A promising future direction may consist in adapting the above techniques to \textit{longer-range} interactions $ \alpha, \beta < 1 $ (a regime known to host peculiar effects, like the emergence of massive Dirac edge modes \cite{Viyuela2015}). Entering this realm is likely to require a full analytical solution of Eq.\,\eqref{eq:generalKappa}, which remains the main technical challenge. Finally, it would be interesting to show via scattering methods persistence of Majoranas in the weakly disordered case, a feature established, e.g., in Ref.\,\cite{Pan2021}. This would however require a much more refined technology, involving heavily the scattering matrix and possibly hinging on Levinson's theorem \cite{ReedSimonIII,GP2013,kellendonk2005}. Inspiration for any attempt in this direction could come from works where scattering theory was successfully applied to disordered systems, see, e.g., the proof of Anderson localization provided in \cite{Ossipov18}.

\emph{Note added in proof.} Recently, a manuscript appeared on Scipost \cite{Bespalov23}, which discussed the decay of edge states in long-range Kitaev chains using the Wiener-Hopf equations, similarly to what is done in \cite{Jaeger20,Jones2022}. The manuscript comments on the difficulty in correctly capturing the leading decay of the edge states within that method. Therefore, the issue of the correct leading-order decay of the Majorana edge modes in the almost finite-range regime remains open, as argued in the present work.

\section*{Acknowledgments}

The authors heartily thank Gian Michele Graf for various discussions and his valuable input on the critical bulk index and the emergence of algebraic decay as involution of exponentials. This research was funded in part by the Swiss National Science Foundation (SNSF) [200021\_207537].  The support of the Deutsche Forschungsgemeinschaft (DFG, German Research Foundation) under Germany’s Excellence Strategy Grant No.\,EXC2181/1-390900948 (the Heidelberg STRUCTURES Excellence Cluster) is also acknowledged.

\appendix

\section{The scattering approach in full generality}
\label{app:GenScattApp}
The aim of this appendix is to clarify how one would operationally employ the scattering approach to study a given (suitable) model. Following the steps below will produce all of the energy-$E$ edge states, provided that the applicability hypotheses of the method are met.
\begin{enumerate}
	\item In a quantum mechanical context, consider a translation invariant Hamiltonian $H$ on position space $ \mathbb{Z} $ or $ \mathbb{R} $. Find its eigenvalues (bands) $ \omega_i (k) $. Let $ \psi_{j,i} (k) $ denote the corresponding eigenvectors. Example: let $ \omega_1 $ be $n_1$-fold degenerate. Then, the corresponding eigenspace is spanned by $ \psi_{j,1} (k) \,, \ j \in \{1,...,n_1\} $.
		
	\item Say the spectrum of $H$ has a gap $ \Gamma \subset \mathbb{R} $. Pick an energy $ E \in \Gamma $. Allow for $ k \in \mathbb{C} $, and solve
	\begin{equation}
		\label{eq:edgeEigen}
		E = \omega_i (k) \,
	\end{equation} 
	for all $i$.
		
	By construction, the equation above cannot have solutions for $ k \in \mathbb{R} $, or else $ \Gamma $ would not be a spectral gap. Label $ k_{l,i} (E) $ the $l$th solution of the $ i $th equation \eqref{eq:edgeEigen} (the set may be empty).
		
	\item Construct the following \textit{scattering state}
	\begin{equation}
		\psi_s (E) \coloneqq \sum_{i,j,l} A_{ijl} \psi_{j,i} (k_{l,i} (E)) \,,
	\end{equation}
	$A_{ijl} \in \mathbb{C}$. Due to translation invariance, the bulk eigenstates can be thought of as plane waves $ \sim \eul^{\mathrm{i} k x} $. By contrast, since $ \mathrm{Im} ( k_{l,i} (E) ) \neq 0 $ by the spectral gap argument above, $ \psi_s (E) $ consists in \textit{evanescent} ($ \eul^{- | \mathrm{Im} (k) | x} $) or \textit{divergent waves} ($ \eul^{+ | \mathrm{Im} (k) | x} $), that vanish (diverge) as $x$ increases. 
		
	Such states can never belong to the bulk Hilbert space, yet they solve the \enquote{formal} eigenvalue problem $ H \psi_{j,i} (k_{l,i} (E)) = E \psi_{j,i} (k_{l,i} (E)) $ given by Eq.\,\eqref{eq:edgeEigen}, as long as the dispersion relations $ \omega_i (k) $ hold even for $ k \in \mathbb{C} $.
		
	\item Consider the wave function $ \psi_s (E; x) $. Restrict the latter to $ \mathbb{N} $ ($ \mathbb{R}_+ $). Impose $ \psi_s (E) \in \hat{\mathcal{H}} $, the edge Hilbert space. This amounts to setting to zero the coefficients $A_{ijl}$ of all \textit{divergent waves}.
		
	\item Impose whatever boundary conditions the problem demands, further fixing the $ A_{ijl} $.
		
	\item The resulting (not identically zero) $ \psi_s (E) $ are bound edge states with energy $ E $, solving
	\begin{equation}
		\hat{H} \psi_s (E) = E \psi_s (E) \,.
	\end{equation}
		
	\item Finally, the number of $A_{ijl}$ coefficients that survived the previous steps represents the number of linearly independent edge states with energy $E$.
\end{enumerate}

\section{Details of long-range case}
\label{app:LRcase}

The purpose of this appendix is to fill the blanks left open in the paragraph devoted to the long-range chain.

As explained, we wish to study edge modes when $ \kappa \to 0 $. This should capture all of the relevant physics when sufficiently close to the critical point ($ \tau $ small). All the same, the decay obtained for the Majorana modes in this approximation is in agreement with various other results in the literature \cite{Jaeger20, Vodola2014, AD17, Vodola2015}. There are hence reasons to presume that the validity of our analysis goes beyond the $ \kappa \to 0 $ region.

The expansions \eqref{eq:EpsDeltaExpansionGeneric}, upon reinstating the complete coefficients, yield
\begin{align}
	\label{eq:EpsDeltaFull}
	\varepsilon_{\kappa} &= \tau - \Gamma(1-\alpha) \cos \left( \frac{\pi}{2} (\alpha -1) \right) \kappa^{\alpha -1} \nonumber \\
	&+ \frac{\zeta (\alpha - 2)}{2} \kappa^2 + \mathcal{O} (\kappa^4) \,, \nonumber \\
	\Delta_{\kappa} &= -\Gamma (1 - \beta) \sin \left( \frac{\pi}{2} (\beta - 1) \right) \kappa^{\beta - 1} + \im \zeta (\beta - 1) \kappa \nonumber \\
	&+ \mathcal{O} (\kappa^3) \,,
\end{align}
where $ \Gamma(\cdot) $ is the extension to complex numbers of the factorial and $ \zeta (\cdot) $ the Riemann Zeta function. Equations \eqref{eq:EpsDeltaFull} are derived from the known series of $ \rm{Li}_{\gamma} (e^{\im z}) \,, z \to 0 $ and $\gamma \in \mathbb{R}$ \cite{L80,L81}.
  
Equations \eqref{eq:EpsDeltaFull} in turn induce the following expression
\begin{align}
	\label{eq:OmegaExp}
	\omega^2 &= \tau^2 - 2 \tau \Gamma (1-\alpha) \cos \left( \frac{\pi}{2} (\alpha-1) \right) \kappa^{\alpha -1} + \tau \zeta(\alpha - 2) \kappa^2 \nonumber \\
	&+ \Gamma^2 (1-\alpha) \cos^2 \left( \frac{\pi}{2} (\alpha-1) \right) \kappa^{2 (\alpha - 1)} + \frac{\zeta^2 (\alpha - 2)}{4} \kappa^4 \nonumber \\
	&+ \Gamma^2 (1 - \beta) \sin^2 \left( \frac{\pi}{2} (\beta-1) \right) \kappa^{2 (\beta - 1)} \nonumber \\
	&- 2 \im \Gamma (1 - \beta) \zeta (\beta - 1) \sin \left( \frac{\pi}{2} (\beta-1) \right) \kappa^\beta \nonumber \\
	&- \zeta^2 (\beta - 1) \kappa^2 + \dots \,, \qquad \kappa \to 0  \,,
\end{align}
again exhibiting the familiar competition between $ \kappa^2, \kappa^{\alpha -1}, \kappa^{2 (\beta - 1)} $. Depending on the winner, we recover the three regimes of the main text: \textit{almost finite-range, hopping-dominated} and \textit{pairing-dominated}.

Before computing inverse scattering lengths and Majorana wave functions, let us dwell on two preliminary results.

First, we notice that $ \cos (\theta_\kappa / 2) $ can be rewritten explicitly in terms of $ \varepsilon_\kappa, \omega_\kappa $ by $ \theta_\kappa = \rm{arctan} (\Delta_\kappa /\varepsilon_\kappa) $ and some goniometry
\begin{equation}
	\label{eq:CosRoot}
	\cos \left( \frac{\theta_\kappa}{2} \right) = \sqrt{\frac{1}{2} + \frac{\varepsilon_\kappa}{2 \omega_k}} \,.
\end{equation}

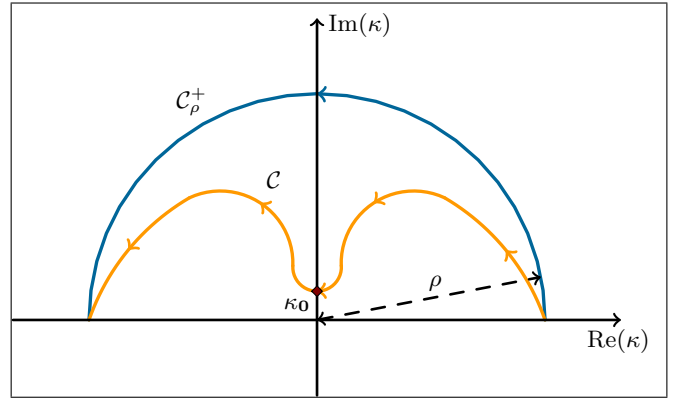
\begin{figure}
	\frame{
	\begin{tikzpicture}[line cap=round,line join=round,x=1cm,y=1cm,
		decoration={%
			markings,
			mark=at position 0.5 with {\arrow[line width=1.25pt]{>}}       
		}]
		\draw [shift={(0,0)},line width=1.25pt,color=qqwwzz,postaction=decorate]  plot[domain=0:3.141592653589793,variable=\t]({1*3*cos(\t r)+0*3*sin(\t r)},{0*3*cos(\t r)+1*3*sin(\t r)});
		\draw [shift={(-0.13758480834375583,-0.9854496604972888)},line width=1.25pt,color=ffzzqq,postaction=decorate]  plot[domain=2.1056232570049334:2.810029734673111,variable=\t]({1*3.0272977657968623*cos(\t r)+0*3.0272977657968623*sin(\t r)},{0*3.0272977657968623*cos(\t r)+1*3.0272977657968623*sin(\t r)});
		\draw [shift={(-1.2734421832044975,0.7544253042986894)},line width=1.25pt,color=ffzzqq,postaction=decorate]  plot[domain=-0.037615018658398114:2.010850611428265,variable=\t]({1*0.9557357598252382*cos(\t r)+0*0.9557357598252382*sin(\t r)},{0*0.9557357598252382*cos(\t r)+1*0.9557357598252382*sin(\t r)});
		\draw [shift={(0.0008532976892332327,0.7004914144351797)},line width=1.25pt,color=ffzzqq,postaction=decorate]  plot[domain=0.061051364412745876:-3.197893780022396,variable=\t]({1*0.31974240023271405*cos(\t r)+0*0.31974240023271405*sin(\t r)},{0*0.31974240023271405*cos(\t r)+1*0.31974240023271405*sin(\t r)});
		\draw [shift={(0.13758480834375417,-0.9854496604972901)},line width=1.25pt,color=ffzzqq,postaction=decorate]  plot[domain=0.3315629189166824:1.0359693965848595,variable=\t]({1*3.027297765796864*cos(\t r)+0*3.027297765796864*sin(\t r)},{0*3.027297765796864*cos(\t r)+1*3.027297765796864*sin(\t r)});
		\draw [shift={(1.2734421832044975,0.7544253042986896)},line width=1.25pt,color=ffzzqq,postaction=decorate]  plot[domain=1.1307420421615282:3.1792076722481917,variable=\t]({1*0.955735759825238*cos(\t r)+0*0.955735759825238*sin(\t r)},{0*0.955735759825238*cos(\t r)+1*0.955735759825238*sin(\t r)});
		\draw [line width=1pt,->] (0,-1)-- (0,4);
		\draw [line width=1pt,->] (-4,0)-- (4,0);
		% Tip of cartesian axes
		%\draw [line width=1pt] (4,0)-- (3.75,0.15);
		%\draw [line width=1pt] (4,0)-- (3.75,-0.15);
		%\draw [line width=1pt] (0,4)-- (0.15,3.75);
		%\draw [line width=1pt] (0,4)-- (-0.15,3.75);
		% end of tips of cartesian axes
		\draw [line width=1pt,dash pattern=on 5pt off 5pt,<->] (0,0)-- (2.945493652675619,0.5692689540521606);
		\draw (0.037170958598680325,4.1882330346646635) node[anchor=north west] {$\mathbf{\mathrm{Im} (\kappa)}$};
		\draw (3.440120886743433,0) node[anchor=north west] {$\mathbf{\mathrm{Re}(\kappa)}$};
		\draw (1.3577186919085844,0.6970217608769462) node[anchor=north west] {$\mathbf{\rho}$};
		\draw (-0.55,0.40) node[anchor=north west] {$\mathbf{\kappa_0}$};
		\draw (-1.9639667603401745,3.192743204938749) node[anchor=north west] {$\mathbf{\mathcal{C}_\rho^+}$};
		\draw (-0.7754738003612607,2.085514720855844) node[anchor=north west] {$\mathbf{\mathcal{C}}$};
		\begin{scriptsize}
			\draw [fill=yqqqqq] (0,0.38075015280379343) ++(-2pt,0 pt) -- ++(2pt,2pt)--++(2pt,-2pt)--++(-2pt,-2pt)--++(-2pt,2pt);
		\end{scriptsize}
	\end{tikzpicture}
}
	\caption{Original $ \mathcal{C}_\rho^+ $ and deformed contour $ \mathcal{C} $. Not represented is the cut of the integrand, which can however be placed so that it avoids intersecting either curve.}
	\label{fig3}
\end{figure}

Secondly, we lay out the solution of a general integral
\begin{equation}
	I(s) = \int_{\mathcal{C}_\rho^+} \kappa^\gamma \eul^{\im \kappa s} \rm{d} \kappa \,,
\end{equation}  
employing the saddle point method in the limit $ s \to + \infty $. Rewrite the integral as
\begin{equation}
	I(s) = \int_{\mathcal{C}_\rho^+} \eul^{s \left(\frac{\gamma}{s} \ln \kappa + \im \kappa \right)} \rm{d} \kappa \,,
\end{equation}  
and assume $ s $ very large. The exponent has a critical point at $ \kappa_0 = \im \gamma / s $. One can deform the original contour $ \mathcal{C}_\rho^+ $ (see Fig. \ref{fig3}) to a new one, $ \mathcal{C} $, such that $ \kappa_0 $ is met as a maximum of $ \mathrm{Re} (\kappa) $. This can be done without changing the value of $ I(s) $, because no singularities are met while deforming. The integral will then almost exclusively depend on the value of the integrand at $ \kappa_0 $, and indeed
\begin{equation}
	\label{eq:SaddlePoint}
	I(s) = - \left( \frac{\im \gamma}{\eul} \right)^\gamma \sqrt{2 \pi \gamma} \frac{1}{s^{\gamma + 1}} + ... \,,
\end{equation}
having applied the standard saddle-point formulas. This result will now be used extensively, with $ \kappa^\gamma \mapsto \kappa^{\alpha - 1}, \kappa^{\beta -1} $.
%\begin{figure}
%	\includegraphics[width=.95\linewidth]{contour.png}
%	\caption{Original $ \mathcal{C}_\rho^+ $ and deformed contour $ \mathcal{C} $. Not represented is the cut of the integrand, which can however be placed so that it avoids intersecting either curve.}
%	\label{fig3}
%\end{figure}

\emph{Inverse localisation lengths.} Recall that we mean, by \enquote{inverse localisation lengths}, the coefficients $ \kappa_j $ appearing at the exponent in Eqs.\,\eqref{eq:LRMajo} or\,\eqref{eq:LRMajoReduced}. When considering edge states of energy $\lambda$ they are, by the scattering approach, solutions of $ \omega (\kappa_j) = \lambda $. The search for zero-energy modes will be conducted by first keeping $ \lambda $ as a small but finite regulator, and eventually sending $ \lambda \to 0 $ to achieve the Majorana limit.
\begin{enumerate}
	\item \textit{Almost finite-range.} Rather than solving $ \omega (\kappa) = \lambda $ directly, we look at its square. Inspection of Eq.\,\eqref{eq:OmegaExp} and selection of the leading orders yield
	\begin{equation}
		\label{eq:DecLengAF}
		\tau^2 + \left( \tau \zeta(\alpha - 2) - \zeta^2 (\beta - 1) \right) \kappa^2 = \lambda^2 \,,
	\end{equation}
	i.e., equivalently $ \kappa^2 = \operatorname{const} $. Exactly two solutions $ \kappa_{1,2} $ are found, and the physics is qualitatively identical to that of the NN case: exponential decay of Majoranas, and at most one Majorana per edge.
	
	\item \textit{Hopping-dominated.} This time, requiring $ \omega^2 (\kappa) = \lambda^2 $ yields
	\begin{equation*}
		\tau^2 - 2 \tau \Gamma(1 - \alpha) \cos \left( \frac{\pi (\alpha - 1)}{2} \right) \kappa^{\alpha - 1} = \lambda^2 \,,
	\end{equation*}
	i.e.,
	\begin{gather}
		\label{eq:AlphaRoot}
		\kappa^{\alpha - 1} = \frac{\tau^2 - \lambda^2}{2 \tau \Gamma(1 - \alpha) \cos (\pi(\alpha - 1)/2)} \,.
	\end{gather}
	The exponent $ \alpha \in \mathbb{R} $ is irrational, unless it lies in the zero-measure set $ \mathbb{Q} \subset \mathbb{R} $. There are hence infinitely many $ (\alpha-1) $-th roots of the constant on the r.h.s. of Eq.\,\eqref{eq:AlphaRoot}. This crucial fact is what will ultimately produce the algebraic decay.
	
	\item \textit{Pairing-dominated.}
	The analysis is analogous to the hopping-dominated case. This time
	\begin{equation*}
		\tau^2 + \Gamma^2 (1 - \beta) \sin^2 \left( \frac{\pi (\beta -1)}{2} \right) \kappa^{2 (\beta - 1)} = \lambda^2 \,,
	\end{equation*}
	entailing
	\begin{equation}
		\label{eq:BetaRoot}
		\kappa^{\beta-1} = \frac{\sqrt{\lambda^2 - \tau^2}}{\Gamma(1 - \beta) \sin (\pi(\beta - 1)/2)}
	\end{equation}
	and the infinite number of solutions is granted by $ \beta $ irrational.
\end{enumerate}

\emph{Decay of Majorana edge modes.} The decay of our model Majorana \eqref{eq:LRMajoReduced} is given by
\begin{equation}
	\label{eq:decay}
	\sum_{\kappa_j} \cos \left( \frac{\theta_{\kappa_j}}{2} \right) \eul^{\im \kappa_j s} \,,
\end{equation}
and can only be estimated upon expansion of the cosine in $ \kappa \to 0 $.

As highlighted in the main text and repeated above, the \textit{almost finite-range} case seems to show no hopes of exhibiting interesting algebraic decay. We therefore avoid treating it altogether, and focus on the \textit{hopping-} or \textit{pairing-dominated} regimes. In either case, there exist infinitely many solutions $ \kappa_j $ of $ \omega (\kappa) = \lambda $, cf.\, Eqs.\,\eqref{eq:AlphaRoot} and \eqref{eq:BetaRoot}, all lying (in first approximation) on a half-circle $ \mathcal{C}_\rho^+ $ of constant radius $ \rho $. Such solutions are moreover homogeneously spaced, and our original sum \eqref{eq:decay} can thus be rewritten as
\begin{equation}
	\label{eq:TruePsi}
	\psi (s) \coloneqq \int_{\mathcal{C}_\rho^+} \cos \left( \frac{\theta_\kappa}{2} \right) \eul^{\im \kappa s} \rm{d} \kappa \,,
\end{equation}
by virtue of the Euler-MacLaurin formula. Equation \eqref{eq:TruePsi} is nothing but the wave function of the Majorana, up to a normalization constant that is yet to be chosen.

It is precisely this normalization that saves us from the divergences seen in the NN case. Being a \enquote{weighted average} of exponential decays $ \eul^{\im \kappa s} $, one can safely claim $ f(s_2) < f(s_1) \,, \ \forall s_2 > s_1 $. Requiring $ \psi(0) $ finite would thus grant, at the very least, $ | \psi(s) | < \infty $ for all $s$. We therefore decide to work with 
\begin{equation}
	\tilde{\psi} (s) \coloneqq \psi (s) / \psi (0) \,,
\end{equation}
where $ \psi(0) $ explicitly reads
\begin{equation}
	\psi (0) = \int_{\mathcal{C}_\rho^+} \cos \left( \frac{\theta_\kappa}{2} \right) \rm{d} \kappa \,.
\end{equation}
If $ \psi (0) $ is not easily evaluated exactly, it is immediate from Eq.\,\eqref{eq:CosRoot} that it diverges when $ \omega (\kappa) \to 0 $. Inserting again our small regulator $ \omega (\kappa_j) = \lambda $, and recalling that $ \omega (\kappa_j) $ is assumed constant on the integration contour leads to the estimate
\begin{equation}
	\label{eq:PsiZero}
	\psi (0) \leq \max_{\kappa \in \mathcal{C}^+_\rho} \sqrt{\frac{\lambda + \varepsilon_\kappa}{2 \lambda}} \cdot \pi \rho \simeq \frac{Q}{\sqrt{\lambda}} \,,
\end{equation}
for some constant $ Q $, as $ \lambda \to 0 $.

As stated, normalizing $ \psi $ formally is impossible unless we evaluate $ \psi (0) $ explicitly. However, Eq.\,\eqref{eq:PsiZero} immediately prompts the idea that divergences of Majorana wave-functions may be cured by substitution
\begin{equation}
	\cos \left( \frac{\theta_\kappa}{2} \right) \mapsto \sqrt{\omega_\kappa} \cos \left( \frac{\theta_\kappa}{2} \right) = \sqrt{\omega_\kappa + \varepsilon_\kappa} \,.
\end{equation}
We thus define our \textit{normalized} Majorana wave-function $ \phi (s) $ as
\begin{equation}
	\phi (s) \coloneqq \int_{\mathcal{C}_\rho^+} \sqrt{\omega_\kappa + \varepsilon_\kappa} \eul^{\im \kappa s} \rm{d} \kappa = \int_{\mathcal{C}_\rho^+} \sqrt{\lambda + \varepsilon_\kappa} \eul^{\im \kappa s} \rm{d} \kappa \,.
\end{equation} 
One can now allow the regulator $ \lambda $ to reach zero. However, some memento of the condition $ \omega_\kappa = 0 $, or equivalently particle-hole symmetry of the scattering state, should be kept. We have learnt in the NN paragraph, cf.\,Eq.\,\eqref{eq:ImposingPHS}, that PHS can be imposed by $ \varepsilon_\kappa - \im \Delta_\kappa = 0 $. Thus
\begin{equation}
	\varepsilon_\kappa = \frac{\varepsilon_\kappa + \varepsilon_\kappa}{2} = \frac{\varepsilon_\kappa + \im \Delta\kappa}{2} \,,
\end{equation}
and
\begin{equation}
	\sqrt{\omega_\kappa} \cos \left( \frac{\theta_\kappa}{2} \right) = \sqrt{\frac{\varepsilon_\kappa + \im \Delta\kappa}{2}} \,.
\end{equation} 
Now and only now can we expand in $ \kappa \to 0 $ using \eqref{eq:EpsDeltaExpansionGeneric}
\begin{align}
	\label{eq:ExpCos}
	\sqrt{\frac{\varepsilon_\kappa + \im \Delta\kappa}{2}} &\simeq \sqrt{\frac{\tau + \varepsilon_\alpha \kappa^{\alpha -1} + \im \delta_\beta \kappa^{\beta - 1} + \im \delta_1 \kappa + ...}{2}} \nonumber \\
	&\simeq \sqrt{\frac{\tau}{2}}
	\begin{cases}
		1 + \frac{2 \delta_1}{\tau} \kappa \,, \\
		1 + \frac{2 \varepsilon_\alpha}{\tau} \kappa^{\alpha - 1} \,, \\
		1 + \frac{2 \delta_\beta}{\tau} \kappa^{\beta - 1} \,,
	\end{cases}
\end{align} 
in the \textit{almost finite-range, hopping-dominated} and \textit{pairing-dominated} cases, respectively.

In the limit $ \tau \to 0 $, where the small-$\kappa$ solutions of $ \omega(\kappa) = 0 $ are granted to be the only relevant ones, we thus have
\begin{align}
	\sqrt{\frac{\varepsilon_\kappa + \im \Delta\kappa}{2}} \propto
	\begin{cases}
		\kappa \,, \\
		\kappa^{\alpha - 1} \,, \\
		\kappa^{\beta - 1} \,,
	\end{cases}
\end{align}
which is Eq.\,\eqref{eq:CosThetaK}.

Plugging this into the normalized Majorana wave-function finally yields
\begin{equation}
	\phi (s) \propto 
	\begin{cases}
		s^{-\alpha} \,, \\
		s^{-\beta} \,,
	\end{cases}
\end{equation} 
in the \textit{hopping-} and \textit{pairing-dominated} regimes, having used Eq.\,\eqref{eq:SaddlePoint} to evaluate $ \phi $. Precise constants can be reinstated by using Eq.\,\eqref{eq:EpsDeltaFull} instead of Eq.\,\eqref{eq:EpsDeltaExpansionGeneric} in Eq.\,\eqref{eq:ExpCos} and plugging the complete Eq.\,\eqref{eq:SaddlePoint} in the last equation we wrote.

Let us close by pointing out that the results above reproduce current literature everywhere but in the almost finite-range regime. There, Refs.\,\cite{Jaeger20,Jones2022} predict algebraic decay $ s^{- \min \{\alpha, \beta\}} $, whereas we observe exponentially localized modes. The discrepancy is not alarming and well-understood. It stems from a known accident: leading orders in position and momentum space often do not match. 
	
The scattering approach crucially needs infinitely many solutions of $ \omega (\kappa) = 0 $ to yield algebraic decay. These are intuitively expected to exist for \textit{any} real value of $ \alpha, \beta $, since the polylogs in Eq.\,\eqref{eq:epsDelta} are always multi-valued with a log-type Riemann surface. Our method is only sensitive to such singularity when it is exhibited by the leading $ \kappa $-order of $ \omega (\kappa) $, namely when $ \kappa^{\alpha-1} $ or $ \kappa^{2 (\beta - 1)} $ dominate the expansion in Eq.\,\eqref{eq:OmegaExp}. This is the case in the hopping- or pairing-dominated regimes, but not in the almost finite-range case. There, our first order approximations become blind to the log-type singularity, and only capture exponential corrections to the actual power-law decay in position space.

The authors are nonetheless certain that the disagreement would be cured if an analytical solutions to $ \omega (\kappa) = 0 $, wished for in the discussion paragraph, was available.

\section{Winding at criticality, details}
\label{app:windCrit}

In this appendix, we show by direct integration that $ w_c = 1/2 $ in the NN case and determine $ w_c $ in the long-range case, for any value of $ \alpha, \beta $, using the \enquote{geometric} reasoning presented in the main text.

In the nearest-neighbor case the direct computation is especially straightforward
\begin{equation}
	\label{eq:bulkIndNew}
	w = - \frac{1}{2 \pi} \oint \mathrm{d} \theta_k \,.
\end{equation}
Without loss of generality we can pick $\mu_{c}=1$ and obtain 
\begin{equation}
	\label{eq:thetaCrit}
	\theta_k^c \coloneqq \theta_k \rvert_{\mu = \mu_c = +1} = \operatorname{arctan} \left( \frac{\sin \ k}{1 - \cos \ k} \right) \,.
\end{equation}
Simple calculus shows that $ \rm{d} \theta_k^c / \rm{d} k = -1/2 $, so that by Eq.\,\eqref{eq:bulkIndNew}
\begin{equation}
	w_c = - \frac{1}{2 \pi} \int_{- \pi}^\pi \rm{d} k  \frac{\rm{d} \theta_k^c}{\rm{d} k} = \frac{1}{2} \,,
\end{equation}
as stated.

The long-range case can now be tackled. By the reasoning in the main text, one predicts $ w_c $ by inspection of $ \varepsilon_k / \Delta_k $ as $ k \to 0 $. Recalling that
\begin{align}
	\varepsilon_{k} \rvert_{\mu = \mu_c} &= \varepsilon_\alpha k^{\alpha -1} + \varepsilon_2 k^2 + \mathcal{O} (k^4) \nonumber \\
	\Delta_{k} \rvert_{\mu = \mu_c} &= \delta_\beta k^{\beta - 1} + \delta_1 k + \mathcal{O} (k^3) \,,
\end{align}
one sees that different leading orders will produce different limits of $ \varepsilon_k / \Delta_k $. More specifically
\begin{equation}
	\left. \frac{\varepsilon_k}{\Delta_k} \right\rvert_{\mu = \mu_c} \simeq 
	\begin{cases}
		(\varepsilon_2/\delta_1) k \,, & \alpha>3 \ \text{and} \ \beta > 2 \\
		(\varepsilon_\alpha/\delta_1) k^{\alpha - 2} \,, & \alpha<3 \ \text{and} \ \beta > 2 \\
		(\varepsilon_2/\delta_\beta) k^{3 - \beta} \,, & \alpha>3 \ \text{and} \ \beta < 2 \\
		(\varepsilon_\alpha/\delta_\beta) k^{\alpha - \beta} \,, & \alpha<3 \ \text{and} \ \beta < 2
	\end{cases}
\end{equation}
as $ k \to 0 $. Inspection of the cases above reveals a simpler structure
\begin{equation}
	\left. \lim_{k \to 0} \frac{\varepsilon_k}{\Delta_k} \right\rvert_{\mu = \mu_c} = 
	\begin{cases}
		0 \,, & \alpha > 2 \ \text{or} \ 1 < \beta < \alpha \\
		\infty \,, & 1 < \alpha < 2 \ \text{and} \ \beta > \alpha \,,
	\end{cases}
\end{equation}
which in turn implies
\begin{equation}
	w_c = 
	\begin{cases}
		\frac{1}{2} \,, & \alpha > 2 \ \text{or} \ 1 < \beta < \alpha \\
		0 \,, & 1 < \alpha < 2 \ \text{and} \ \beta > \alpha \,.
	\end{cases}
\end{equation}

\section{Remarks on particle-hole symmetry and quasi-particle interpretation}
\label{app:Interpretation}

The goal of the following paragraphs is formalizing our notion of particle-hole symmetry, introducing the operation of particle-hole conjugation and reporting an interpretation of the quasi-particle picture induced by the BdG structure and successive diagonalization, cf.\,Eq.\,\eqref{eq:hamDiag}. Such clarifications are inserted in the interest of readability, but they are not new. Identical information can be found elsewhere in the literature, see, e.g., Ref.\,\cite{Chiu16}.

We say a Hamiltonian $H$ is \textit{particle-hole symmetric} if the \enquote{mathematical tautology} \cite{ZirnbauerPHS}
\begin{equation}
	\label{eq:partHoleSymm}
	\mathcal{P} H \mathcal{P}^{-1} = -H 
\end{equation}
is satisfied, where $ \mathcal{P} $ denotes the antilinear operation of particle-hole conjugation, acting on the fermionic creation and annihilation operators in position space as
\begin{equation}
	\label{eq:partHoleConj}
	\phs (\lambda c_i) \phs^{-1} = \bar{\lambda} c_i^\dagger \,, \qquad \lambda \in \mathbb{C} \,.
\end{equation}
Written as in Eq.\,\eqref{eq:partHoleConj}, $ \phs $ looks very much like hermitian conjugation. The two are however different, as is immediately seen by applying them to the quadratic and hermitian operator $ O = c_i c^{\dagger}_j + c_j c^\dagger_i $. Indeed, $ O^{\dagger} $ is equal to itself. However, Eq.\,\eqref{eq:partHoleConj} equivalently means $ \phs \lambda c_i = \bar{\lambda} c_i^\dagger \phs$, so that (assuming $i \neq j$)
\begin{equation}
	\phs O \phs^{-1} = (c^\dagger_i c_j + c^\dagger_j c_i) = - O \,,
\end{equation} 
where the last member is found by applying the fermionic canonical anticommutation relations.	

Some people refer to Eq.\,\eqref{eq:partHoleSymm} as a mathematical tautology because, as can be seen by the simple example of $O$, any quadratic fermionic operator enjoys this property, i.e., should be deemed particle-hole symmetric. Setting such debates aside, it is now apparent that any BdG (quadratic) Hamiltonian is particle-hole symmetric according to the definition above.

One can of course wonder what this implies for the momentum space BdG matrix, namely $ H(k) $ in Eq.\,\eqref{eq:hamk}. Applying $\phs$ to $ H $ as prescribed induces a \enquote{momentum-space} particle-hole conjugation, which explicitly reads
\begin{equation}
	H(k) \mapsto (\sigma_x \mathcal{K}) H(k) (\sigma_x \mathcal{K})^{-1} \,,
\end{equation}
where $ \mathcal{K} $ denotes complex conjugation and $ \sigma_x $ is the first $ 2 \times 2 $ Pauli matrix. The Hamiltonian is then called particle-hole symmetric if it satisfies the momentum-space version of Eq.\,\eqref{eq:partHoleSymm}, namely
\begin{equation}
	(\sigma_x \mathcal{K}) H(k) (\sigma_x \mathcal{K})^{-1} = - H(-k) \,.
\end{equation}

Now, seen as an operation on the quasi-particles $ \gamma_k $ of Eq.\,\eqref{eq:bgTrans}, particle-hole conjugation on $H$ is embodied by the map $ \tilde{\phs} (\cdot) \tilde{\phs}^{-1} $ acting like
\begin{equation}
	\gamma_k \mapsto \tilde{\phs} (\gamma_k) \tilde{\phs}^{-1} = \gamma_{-k}^\dagger \,.
\end{equation}
This favours the following interpretation of the Bogoliubov modes. 

Start with $ \gamma^\dagger_k $. This is a mode with positive energy $ \omega_k $ and positive momentum $ k > 0 $. We see this operator as \enquote{creation of a particle}. Under $ \tilde{\phs} $, the latter is mapped to $ \gamma_{-k} $, a negative-energy mode with negative momentum. If anti-particles have opposite energy with respect to their particle counterpart, it is appealing to interpret this as \enquote{creation of an anti-particle}. By the same token, one can view $ \gamma_k $ as \enquote{annihilation of a particle} and $ \gamma^{\dagger}_{-k} $ as \enquote{annihilation of an antiparticle}.

Seeing the problem from this angle justifies our notation: $\chi$ ($\varphi$) referred to particle (antiparticle) states, whereas the subscript $ (\cdot)_+ $ ($ (\cdot)_- $) to positive (negative) energy. The appeal of this interpretation is actually twofold, as it also provides an intuitive explanation of why $ \chi_+ $ and $ \varphi_- $ collapse to the same Majorana: being a particle-antiparticle pair, they become indistinguishable when their associated excitation energy approaches zero.

\section{Nearest neighbor problem with finite-difference equations}

The aim of this section is to review how finite-difference methods allow for the detection of Majoranas in the nearest neighbor case. This is how the problem was solved in Kitaev's seminal paper \cite{KitaevChain}. It will be seen that the end result is in perfect agreement with Eq.\,\eqref{eq:finalMajo}.

Kitaev adopts conventions that differ slightly from ours. His nearest neighbor Hamiltonian indeed reads
\begin{equation}
	\label{eq:hamKit}
	H= \sum_{t=1}^N - g (c^{\dagger}_{t} c_{t} - \frac{1}{2}) - j (c^{\dagger}_{t} c_{t+1} + \rm{h.c.} ) - \Delta ( c^{\dagger}_{t} c^{\dagger}_{t+1}+ \rm{h.c.}) \,,
\end{equation}
whereas ours is, cf.\,\eqref{eq:ham},
\begin{equation}
	\label{eq:hamNN}
	H= \sum_{t=1}^N \mu (2 c^{\dagger}_{t} c_{t} - 1) - (c^{\dagger}_{t} c_{t+1} + \rm{h.c.}) - (c^{\dagger}_{t} c^{\dagger}_{t+1} + \rm{h.c.}) \,.
\end{equation}
We notice that Eq.\,\eqref{eq:hamNN} is obtained from Eq.\,\eqref{eq:hamKit} by $ g \mapsto -2 \mu$, $ j \mapsto 1 $, $ \Delta \mapsto 1 $. In order to make contact with existing literature, we will solve the finite-difference problem adopting the conventions of Ref.\,\cite{KitaevChain}, and later show that Eq.\,\eqref{eq:finalMajo} is recovered by the substitutions above.

Start by writing Eq.\,\eqref{eq:hamKit} in BdG form
\begin{align}
	\label{eq:BdGKit}
	H = &\sum_t \left[ - \frac{g}{2} (c^\dagger_t c_t - c_t c^\dagger_t) -\frac{j}{2} ( c^\dagger_t c_{t+1} - c_{t+1} c^\dagger_t \right. \nonumber \\
	&\left. + c^\dagger_{t+1} c_t - c_t c^\dagger_{t+1} ) + \frac{\Delta}{2} (c_t c_{t+1} - c_{t+1} c_t \right. \nonumber \\ 
	&\left. + c^\dagger_{t+1} c_t^\dagger - c^\dagger_t c^\dagger_{t+1}) \right].
\end{align}
The $ 2N \times 2N $ matrix $ H_{BdG} $ is read off from here.

Consider now a $2N$-vector $\psi$, candidate eigenvector of $H_{BdG}$. Mimicking the BdG doubling of dimensions, we write it as the juxtaposition of two vectors $ \chi $ and $ \varphi $. More precisely, $ \psi_m =  \chi_m \,, \ 1 \leq m \leq N $ and $ \psi_{m+N} = \varphi_m \,, \ 1 \leq m \leq N $, where $ (\cdot)_m $ denotes the $m$th entry of a vector. $\psi$ is then an eigenvector of $ H_{BdG} $ with eigenvalue $E$ if $ \chi $ and $ \varphi $ satisfy the system of coupled finite difference equations
\begin{equation}
	\label{eq:finiteDiff}
	\begin{cases}
		E \chi_m = -j (\chi_{m-1} + \chi_{m+1}) - g \chi_m + \Delta (\varphi_{m-1} - \varphi_{m+1}) \\
		E \varphi_m = j (\varphi_{m-1} + \varphi_{m+1}) + g \varphi_m - \Delta (\chi_{m-1} - \chi_{m+1}) \,. \\
	\end{cases}
\end{equation}
This is in general not easy to solve: one can decouple the two equations at the price of turning an order-2 into an order-4 problem. However, when looking for Majoranas, PHS shall be imposed. This is tantamount to requiring $ \chi_m = \pm \varphi_m $. Moreover, $E=0$. 

Pick then the first option $ \chi_m = \varphi_m $. The system \eqref{eq:finiteDiff} reduces to two identical equations in $ \chi $ or $ \varphi $. This is indeed akin to the \enquote{collapse} of particle $ \chi_{\pm} $ and antiparticle $ \varphi_\mp $ solutions of the scattering approach, see above Eq.\,\eqref{eq:ChiMajo}. The equation in $ \chi $ explicitly reads
\begin{equation}
	\label{eq:finiteDiffMajo}
	(\Delta + j) \chi_{m+2} - g \chi_{m+1} - (\Delta - j) \chi_m = 0 \,.
\end{equation}
Equivalently, denoting by $L$ the left shift $ (L \psi)_m = \psi_{m+1} $, Eq.\,\eqref{eq:finiteDiffMajo} has operator form
\begin{equation}
	\label{eq:OpFinDiff}
	((\Delta+j)L^2 + g L - (\Delta - j) ) \chi = 0 \,.
\end{equation}
The general solution of equations like Eq.\,\eqref{eq:OpFinDiff} is known (see, e.g., Ref.\,\cite{FiniteDiffBook}), and reads
\begin{equation}
	\label{eq:ChiM}
	\chi_m = (A \lambda_1^m + B \lambda_2^m) \,,
\end{equation}
where $ A,B \in \mathbb{C} $, and $ \lambda_{1,2} $ are the two roots of
\begin{equation}
	(\Delta+j) \lambda^2 + g \lambda - (\Delta - j) = 0 \,,
\end{equation}
namely
\begin{equation}
	\label{eq:Lambdas}
	\lambda_{1,2} = \frac{-g \pm \sqrt{g^2 + 4 \Delta^2 - 4 j^2}}{2 (\Delta + j)} \,.
\end{equation}

The eigenvector $\psi$ is put in one-to-one correspondence with an eigenmode (which we again call $\psi$) by the map
\begin{equation}
	\psi \mapsto \sum_{t=1}^N \chi_t c_t + \varphi_t c^\dagger_t \,.
\end{equation}
Combining $ \chi_m = \varphi_m $, Eqs.\,\eqref{eq:ChiM} and \eqref{eq:Lambdas}, one therefore concludes
\begin{equation}
	\psi = \sum_{t=1}^N (A \lambda_1^t + B \lambda_2^t) (c_t + c_t^\dagger) \,.
\end{equation}

We can now make contact with our own results. Set $ g = - 2 \mu \,, \ j=\Delta=1 $ in Eq.\,\eqref{eq:Lambdas}:
\begin{equation}
	\lambda_1 = 0 \,, \qquad \lambda_2 = -(g/2) = \mu \,,
\end{equation}
so that
\begin{equation}
	\label{eq:MajoKit}
	\psi = A \sum_{t=1}^N \mu^t (c_t + c_t^\dagger) \,.
\end{equation}
This is precisely $ \chi_+^{(1)} (\hat{\kappa}_1) $, cf.\,Eq.\,\eqref{eq:MajoChi}.  The antisymmetric combination of $ c_t, c_t^\dagger $ will similarly be obtained by imposing $ \chi_m = - \varphi_m $, and will coincide with Eq.\,\eqref{eq:ChiMajo}. Taking a linear combination of the two yields precisely Eq.\,\eqref{eq:finalMajo}. The two approaches are therefore seen to be equivalent.

\bibliography{bib}

\end{document}